\documentclass[lettersize,journal]{IEEEtran}
\usepackage{epsfig,rotating,setspace,latexsym,amsmath,epsf,amssymb,amsfonts,bm,theorem,booktabs,subfigure,epstopdf}
\usepackage{cite,authblk,hyperref}
\usepackage{bbm}

\usepackage{algorithm}
\usepackage{algpseudocode}%

\usepackage{color}

\definecolor{commentcolor}{RGB}{110,154,155}   

\IEEEoverridecommandlockouts
\allowdisplaybreaks

\newcommand{\rev}[1]{{\color{black}#1}}

\begin{document}
	\bstctlcite{IEEEexample:BSTcontrol} 
	
	\title{Self-Supervised RF Signal Representation Learning for NextG Signal Classification with Deep Learning}
	
	\author[1]{Kemal Davaslioglu}
	\author[1]{Serdar Bozta\c{s}}
	\author[1]{Mehmet Can Ertem}
	\author[2]{Yalin E. Sagduyu}
	\author[3]{Ender Ayanoglu}
	\affil[1]{\normalsize University Technical Services, Inc. (UTS),  Greenbelt, MD, USA.} 
\affil[2]{\normalsize Virginia Tech, National Security Institute, Arlington, VA, USA. 
}
\affil[3]{\normalsize 
	Center for Pervasive Communications and Computing (CPCC), University of California, Irvine, Irvine CA, USA. 
}
\maketitle

\begin{abstract}
	Deep learning (DL) finds rich applications in the  wireless domain to improve spectrum awareness. Typically, DL models are either randomly initialized following a statistical distribution or pretrained on tasks from other domains in the form of transfer learning without accounting for the unique characteristics of wireless signals. Self-supervised learning (SSL) enables the learning of useful representations from Radio Frequency (RF) signals themselves even when only limited training data samples with labels are available. We present a self-supervised RF signal representation learning method and apply it to the automatic modulation recognition (AMR) task by specifically formulating a set of transformations to capture the wireless signal characteristics. We show that the sample efficiency (the number of labeled samples needed to achieve a certain performance) of AMR can be significantly increased (almost an order of magnitude) by learning signal representations with SSL. This translates to substantial time and cost savings. Furthermore, SSL increases the model accuracy compared to the state-of-the-art DL methods and maintains high accuracy when limited training data is available.
\end{abstract}


\section{Introduction}\label{sec:Introduction}
\emph{Deep learning} (DL) finds diverse applications in wireless communications such as in wireless signal classification, spectrum sensing, spectrum sharing, resource allocation and provisioning, and attack detection and mitigation \cite{erpek2020deep}. 
In particular, DL has emerged as a key enabler for \emph{next-generation (NextG) communications}
by providing novel means to solve complex tasks \cite{Ayanoglu22}.
One particular task that benefits from DL is \emph{wireless signal classification} that serves a variety of applications ranging from user equipment (UE) identification and PHY-layer authentication to jammer detection. To that end, \emph{automatic modulation recognition} (AMR) with DL has drawn major attention to showcase the use of deep neural networks (DNNs) in capturing the intrinsic characteristics of wireless signals especially in the low signal-to-noise ratio (SNR) regime, where statistical and conventional machine learning methods with hand-crafted features (such as high-order cumulants as in \cite{Stella21}) often fail \cite{dyspan19, milcom19, kim21, Stella21}. 

Despite the unique performance promises, DL in Radio Frequency (RF) applications faces several challenges. First, the received signal is subject to noise, channel, traffic, and interference effects. Modeling or estimating these effects individually \rev{or determining which signal features to use and how to preprocess them is a very difficult task. For example, a recent study in \cite{Stella21} transformed the complex-valued signal waveforms into images based on the density of data features to express its statistical information.} 
Second, collecting representative RF data is generally an expensive and time consuming endeavor. An example is dynamic spectrum sharing (DSA) as in the Citizens Broadband Radio Service (CBRS) band, where the 5G communications need to share the spectrum with incumbent users (such as radar) by classifying the received signals and opportunistically accessing the spectrum when idle. Thus, it is essential for wireless systems to train classifiers even when limited training data samples are available.   

Currently, a popular practice in using the DL models for spectrum classification tasks (e.g., spectrum sensing and waveform identification) is to employ \emph{transfer learning} where the DNN weights are initialized based on a computer vision task trained using a vision dataset \cite{kim21} or they are randomly initialized \cite{RML201610a,MCLDNN,amr_benchmark}.
However, these DL models do not exploit the unique characteristics of RF spectrum data (e.g., channel distortions, phase and frequency offsets, and noise effects) as they do not use the signal representations during pretraining. 

\emph{Self-supervised representation learning} from unlabeled data has recently attracted major attention in the natural language processing (NLP) and computer vision areas \cite{rotnet,word2vec,SimCLRv2,mocov1,mocov2,mocov3}.
The representations learned using self-supervised learning (SSL) have been shown to outperform their supervised counterparts \cite{mocov1}. One of the main goals in SSL is to learn representations and pretrain models that can be transferred to \emph{downstream tasks} by fine-tuning. In computer vision, examples of downstream tasks are detection and segmentation. In wireless communication applications, the downstream tasks may include determining the presence of a signal and identifying its emitter, protocol, and modulation. 
In radar applications, the downstream tasks may include resolving Pulse Descriptor Words (PDWs) such as pulse repetition patterns, pulse width, SNR, duty cycle, and time, frequency and direction of arrival. 

SSL methods typically require \emph{a pretext task} to learn from and \emph{a loss function} to optimize \cite{mocov1}. The pretext task is solved to learn a good representation of the data, but typically is not the true purpose of learning. For example, predicting the rotation of an image \cite{rotnet} or predicting the center word given the surrounding words \cite{word2vec} are some commonly used pretexts. \textcolor{black}{To the best of our knowledge, there are only two prior studies that have applied SSL to AMR task \cite{SSL_Army,SSL_Yun}. Both of these references are based on the SimCLR method \cite{SimCLR}, in which  \cite{SSL_Yun} uses adding Gaussian noise and carrier frequency offset (CFO) for augmentation, while \cite{SSL_Army} uses rotations. In comparison to these studies, we present an SSL method that is based on the MoCo-v3 framework. In addition, we include five AMR-related transformations for the data augmentation step. In Section~\ref{sec:performance}, we compare the performance of our approach with \cite{SSL_Army,SSL_Yun} and demonstrate substantial gains.}

In this letter, we study AMR as an example of wireless signal classification using DL. Our contributions are twofold: (i) \rev{We propose an efficient self-supervised signal representation method for AMR to learn the signal representations in RF applications and reduce the need for labeled data}. 
(ii) We propose a set of data augmentation transformations that do not alter the semantic information of the data. These transformations may significantly change the amplitude or phase of the signal, but they preserve the semantic meaning (e.g., modulation type, radar signal type, and emitter information).
The DL model is pretrained with contrastive learning using the proposed transformations \emph{without the need for any labels}. Then, the AMR task uses the backbone of the pretrained model and fine-tunes its weights using labeled data. The self-supervised pretraining significantly improves the sample efficiency of the learning process, which is defined as the number of labeled samples required to achieve a certain performance (in our case, classification accuracy). 
Furthermore, SSL increases the accuracy of the DL model and surpasses the performance of the existing supervised learning benchmarks.

The remainder of the letter is organized as follows: Section~\ref{sec:SystemModel} defines the AMR problem and data augmentations. Section~\ref{sec:ssl} describes the SSL approach. Section~\ref{sec:performance} evaluates the performance and   Section~\ref{sec:Conclusion} concludes the letter.

\section{System Model} \label{sec:SystemModel}
Suppose that a receiver receives the following signal with a single antenna and over a single channel:
\begin{align}
	y(i) = A(i) e^{j(\omega i+\phi)} x(i) + n(i)
\end{align}
for $i=1,\cdots,N$, where $x(i)$, $y(i)$, and $n(i)$ are the $i$th transmitted signal, received signal, and noise samples, respectively,
$A(i)$ is the channel gain, $\omega$ is the frequency offset, and $\phi$ is the phase offset. The received signal can be expressed in a vector form $\textbf{y}=[y_1,\cdots,y_N]$, where all $N$ signal samples are modulated using the same modulation.

The goal of the AMR classifier is to train a model that learns to map the received signal sample $\textbf{y}$ to a modulation class, that is, $f_{\boldsymbol{\theta}}(\textbf{y}): \mathbb{R}^d \rightarrow \mathbb{R}^c$, where parameters (weights and biases) of the neural network are represented by $\boldsymbol{\theta}$. The dimension of the received signal is $d$ and there are $c$ classes (labels) such that when a digitized RF waveform $\textbf{y}$ is received and input to the modulation recognition model, a label is returned, i.e., $f_{\boldsymbol{\theta}}(\textbf{y}) \in \mathbb{R}^{c}$. This problem can be formulated as a multiclass classification problem which can be solved by minimizing the cross-entropy loss that is given by 
\begin{align}
	\ell_{CE}(\textbf{y}) = - \sum_{i=0}^{c} \beta_i \log \left(  p_{i} \right),
\end{align}
where $\beta_i$ is a binary indicator that specifies if the vector $\textbf{y}$ belongs to class $i$ or not, that is, $\beta_i=1$ if $\textbf{y}$ belongs to class $i$ and is $0$ otherwise, and $p_i$ is  the output of the neural network that denotes the predicted probability. The cross-entropy loss is minimized to train the DNN with supervision. 

\begin{figure}[!tb]
	\centering
	\includegraphics[width=0.91\columnwidth]{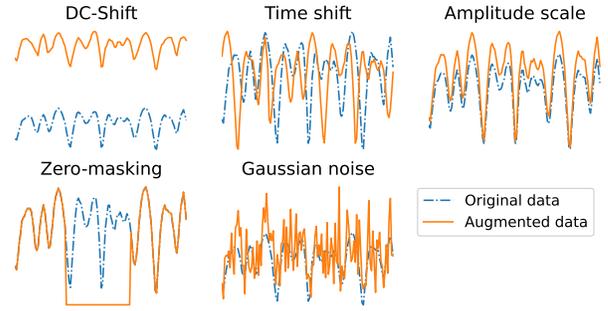}	
	\caption{RF waveform augmentations.}
	\label{fig:rf_augmentation}
\end{figure}

\begin{table}[tb!]
	\small
	\centering
	\caption{Parameter ranges used to generate waveform transformations.
	}
	\label{tab:data_aug_ranges}
	\begin{tabular}{c|cc}
		\hline
		Transformation & Min & Max \\ \hline
		DC shift & 0 & 0.0001 \\
		Time shift (samples) & -40 & 40 \\
		Amplitude scale & 0.8 & 1.2 \\
		Zero-masking (samples) & 0 & 25\\
		Additive White Gaussian noise & \multicolumn{2}{c}{$\mathcal{N}(0,10^{-5})$}  \\ \hline
	\end{tabular}
\end{table}

\subsection{Data augmentation methods}
A key component of \emph{contrastive learning} is a set of \emph{data augmentation methods} that do not alter the semantic information of the data. Most contrastive learning algorithms learn representations that are maximally similar for augmented instances of the same sample and minimally similar for those from different samples. In computer vision, commonly used transformations include random resized cropping, horizontal flipping, color jittering, grayscale conversion, blurring, and solarization \cite{mocov3}. However, most of these transformations cannot be applied to the RF signals. In this letter, we carefully select five transformations to augment the RF signals during the self-supervised training stage while maintaining the semantic information of the signals. These transformations are adding a DC shift, shifting the signal in time domain (time shift), amplitude scaling (multiplying with a constant factor), zero-masking (nulling a set of consecutive samples), and adding additive white Gaussian noise (AWGN). Although these transformations can significantly alter the  numerical values, they preserve the semantic information of the data (e.g., constellation map regions and frame structure). Fig.~\ref{fig:rf_augmentation} illustrates an example of each transformation where the original signal and its augmented version are shown together. The x-axis denotes the time and y-axis denotes the amplitude of signal. Table~\ref{tab:data_aug_ranges} presents the parameter ranges of these transformations. Each transformation is applied with a random value within the specified range, which depend on the data. For example, as the values in the RML2016.10a \cite{RML201610a} dataset are very small, we add a zero-mean noise with a small variance.
 
\begin{figure}[tb!]
	\centering
	\includegraphics[width=\columnwidth]{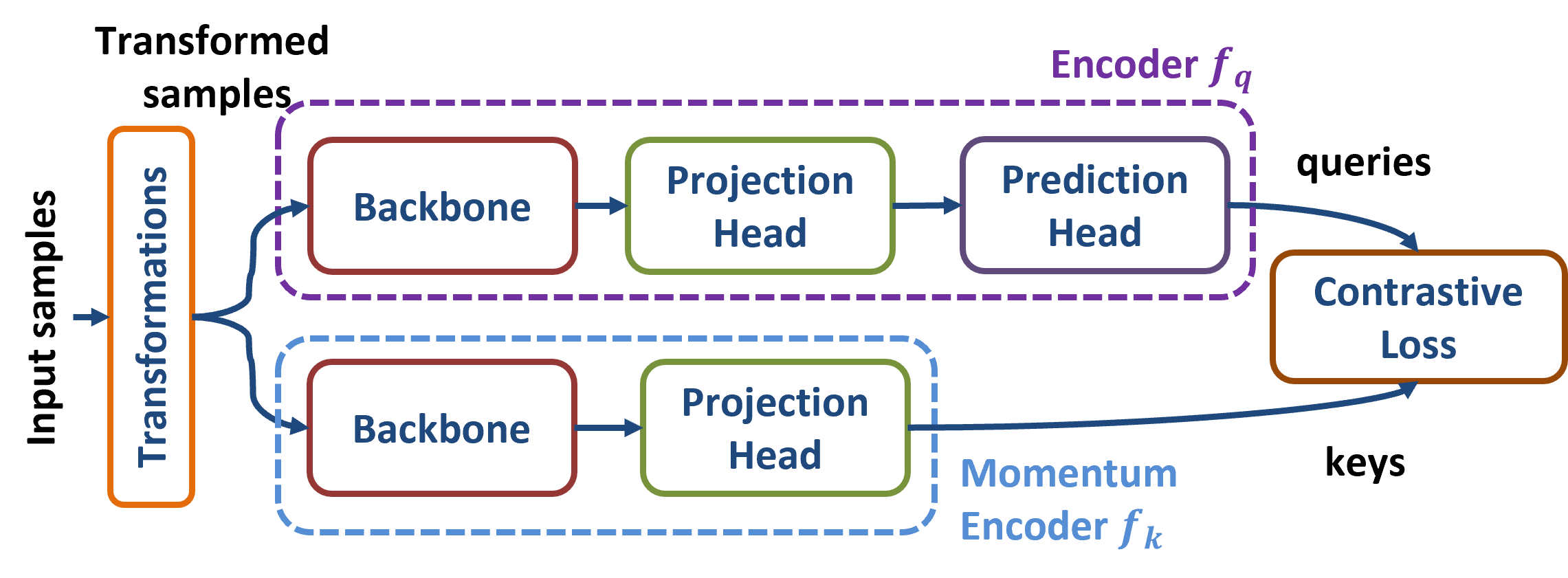}
	\caption{Contrastive learning using MoCo-v3.} 
	\label{fig:moco-v3}
\end{figure}

\section{Self-Supervised Learning} \label{sec:ssl}
Contrastive methods learn feature representations by enforcing similar features to be close to each other (\emph{positive samples}) and dissimilar features to be far away from each other (\emph{negative samples}). Fig.~\ref{fig:moco-v3} shows the MoCo-v3 framework \cite{mocov3} that uses two encoders and the pseudocode of the MoCo-v3 algorithm is presented in Algorithm~\ref{algo:mocov3}. 
The first encoder $\textbf{f}_q$ includes a backbone (e.g., ResNet50 model \cite{resnet50}), a projection head, and a prediction head. The backbone consists of a convolutional neural network (CNN), whereas the projection and prediction heads are \rev{made up of} multi-layer perceptrons (MLPs), i.e., feedforward neural networks. The projection and prediction heads consist of 3-layer and 2-layer MLPs, respectively, where each linear layer (except the last one) is followed by a batch normalization and a Rectifying Linear Unit (ReLU) activation. The second encoder $\textbf{f}_k$ only has a backbone and a projection head. We denote the output vectors of the encoders $\textbf{f}_q$ and $\textbf{f}_k$ as $\textbf{q}$ and $\textbf{k}$, respectively, both have 256 dimensions. MoCo-v3 uses the Information Noise-Contrastive Estimation (InfoNCE) loss \cite{info_nce} that is given by 
\begin{align} \label{eqn:infonce}
	\ell_{\textbf{q}} = - \log \left( \frac{\exp\left(\textbf{q} \textbf{k}_{+}/\tau\right)}{\exp\left(\textbf{q} \textbf{k}_{+}/\tau\right) + \sum_{\textbf{k}_{-}} \exp\left(\textbf{q} \textbf{k}_{-}/\tau\right)} \right),
\end{align}
where $\tau$ is a temperature hyper-parameter. \rev{Each RF waveform in a batch is augmented twice using the RF transformation function $A(\cdot)$ that applies all the data transformations discussed in Table~\ref{tab:data_aug_ranges} consecutively. The output vectors of the same waveform form a positive pair, denoted by $\textbf{k}_{+}$, and others form the negative pairs, denoted by $\textbf{k}_{-}$. In each batch, we calculate the InfoNCE loss in (\ref{eqn:infonce}) and use the queries $\textbf{q}$ to \rev{retrieve} the corresponding keys $\textbf{k}$. The backpropagation loss is calculated and the query encoder $\textbf{f}_q$ is updated. The second encoder $\textbf{f}_k$ is updated by the moving average of the query encoder $\textbf{f}_q$ to maintain consistency.}

\begin{algorithm}[tb!]
	\caption{MoCo-v3 algorithm for SSL.}\label{algo:mocov3}
	\label{table:mocov3_algo}
	\small
	\begin{algorithmic}[1]
		\Procedure{MoCo-v3}{$x$} 
		\State{\textbf{Input:} Initial \textcolor{black}{momentum encoder} $\textbf{f}_k$, initial query encoder $\textbf{f}_q$, RF transformation function $A(\cdot)$, momentum update coefficient $\alpha$, and temperature coefficient $\tau$.}
		\For{\textcolor{black}{$\textbf{x} \in \mathcal{X}$}} \Comment{\textcolor{black}{Load a batch of $N$ samples}}
		\State $\textbf{x}_1=A(\textbf{x})$ and $\textbf{x}_2 = A(\textbf{x})$
		\Comment{Augment $\textbf{x}$ twice.}
		\State $\textbf{q}_1 = f_q(\textbf{x}_1)$ and $\textbf{q}_2 = f_q(\textbf{x}_2)$ \Comment{Forward pass $\textbf{x}_1$ and $\textbf{x}_2$ through the query encoder.}
		\State $\textbf{k}_1 = f_k(\textbf{x}_1)$ and $\textbf{k}_2 = f_k(\textbf{x}_2)$ \Comment{Forward pass $\textbf{x}_1$ and $\textbf{x}_2$ through \textcolor{black}{the momentum encoder}.}
		
		\State Calculate the InfoNCE loss \rev{in (\ref{eqn:infonce})}. 
		\State Back-propagate the loss. 
		\State Update the query encoder $\textbf{f}_q$. 
		\State Update the \textcolor{black}{momentum encoder} $\textbf{f}_k$ using the momentum by $\textbf{f}_k = \alpha \textbf{f}_k + (1-\alpha) \textbf{f}_q$.
		\EndFor
		
		\EndProcedure		
		
	\end{algorithmic}
\end{algorithm}

In comparison with MoCo-v1 \cite{mocov1} and MoCo-v2 \cite{mocov2}, MoCo-v3 algorithm \cite{mocov3} includes slight changes to provide simplicity, accuracy, and stability improvements. It does not require the use of an external memory. Instead, it uses the encoded features in the same batch. MoCo-v3 has an architectural change where the query encoder $\textbf{f}_q$ uses an extra prediction head. \rev{The MoCo-v3 algorithm was originally proposed for the computer vision domain \cite{mocov3}. Our proposed algorithm builds on the MoCo-v3 framework and extends it to RF applications with two subtle differences}. 
First, the set of transformations presented in this letter for data augmentation are different. Second, the neural network architectures of the query and key encoders in this letter are slightly different (see Section~\ref{sec:performance} for details). 

Overall, AMR model training consists of two steps. In the first step, the signal representations are learned using the MoCo-v3 algorithm. This step is called as \emph{pretraining}. In the second step, the backbone of the encoder is used to initialize the model. Its last layer is changed to have consistent size with the AMR problem. The model is trained end-to-end until convergence, where the weights and biases of the backbone are also updated. This step is called as \emph{end-to-end fine-tuning}. Note that there is also an alternate approach called \emph{linear probing} \cite{mocov3}, where the parameters of the backbone are frozen and only the last layers of the classifier are updated. In this letter, results with linear probing are omitted since its performance is consistently lower than end-to-end fine-tuning. 

\section{Performance Evaluation for AMR} \label{sec:performance}
Performance is evaluated using the RML2016.10a dataset \cite{RML201610a}. This dataset includes 11 modulations (classes). Signals in the dataset cover an SNR range from -20~dB to 18~dB in 2~dB increments. At each SNR, there are 1000 samples per modulation type. The dataset is not readily split into training and test subsets and we observe that the performance significantly depends on the way data is split. In all simulations, we use the same training, validation, and test datasets for repeatability. We leave 20\% of the data  for test and the model is not evaluated on the test dataset during training. For the SSL stage, the model is trained on the remaining 80\% of the data without any labels. For the supervised learning stage, we vary the training set portion of the training and validation sets (constituting 80\% data) from 0.1\% to 90\%, while the remaining data is used for validation. After the training stage, all models are evaluated on the test dataset using the model with the best validation loss. The algorithms are implemented in Python using the PyTorch framework. \rev{The experiments are conducted using a server with an NVIDIA GeForce RTX 3090 Ti Graphics Processing Unit (GPU)}. For the SSL code, we modified the code released by Facebook \cite{mocov3}. We used the temperature parameter as $\tau = 1.0$ as suggested in \cite{mocov3}. \rev{The models are trained for 100 epochs during the SSL stage}. 

Our AMR model modifies the \rev{ResNet50 architecture} \cite{resnet50} to be compatible with the RF data. The first layer of convolutions \rev{is} changed from 3 channels to 1 channel. \rev{We kept the kernel size as (7, 7) and stride size as (2, 2) as in the original model.} The output layer is changed to have 11 neurons representing one for each label. The RF signals \rev{consist} of 128 symbols in the RML2016.10a dataset \cite{RML201610a}. We reshape the in-phase and quadrature components (I/Q) of the RF signal to $1 \times 2 \times 128$ using the channel-first notation. AdamW \cite{adamw} optimizer is used with a learning rate of 0.01. The learning rate in the supervised learning is reduced to its half if the validation loss does not improve for five consecutive epochs. All simulations are run for 500 epochs. If the validation loss does not improve for 20 epochs, the simulations are early terminated.

For comparison, we consider two supervised benchmarks using the same ResNet50 model. The first model uses transfer learning where the model is pretrained on the \rev{ImageNet dataset \cite{imagenet_dataset}}. The second one uses the Xavier initialization that draws from a uniform distribution in $[-1,1]$ and then scales the weights by $\sqrt{6/(m_i+m_{i+1})}$, where $m_i$ is the number of neurons at layer $i$. We also include the Multi-Channel Convolutional Long short-term Deep Neural Network (MCLDNN) \cite{MCLDNN} and 
Parameter Estimation and Transformation
based CNN-GRU Deep Neural Network (PET-CGDNN) models \cite{amr_benchmark} in our comparison. These models currently achieve the best performance reported in the literature for this dataset. \rev{In addition, we implemented two SSL methods based on the SimCLR method for comparison, namely, SimCLR-AWGN \cite{SSL_Yun} that uses the AWGN transformation and SimCLR-5TX that uses the five transformations considered in this letter.}

\begin{table}[tb!]
	\small
	\centering
	\caption{The AMR accuracy for different backbone initializations.}
	\label{tab:amr_comprasion}
	\resizebox{\columnwidth}{!}{
		\begin{tabular}{c|ccccccc}
			\hline
			& \multicolumn{7}{c}{Percent data used in training} \\ \hline
			Method & 0.5\% & 1\% & 5\%    & 10\%   & 50\% &  75\% & 90\%
			\\ \hline
			Xavier init. & 9.1 & 11.4 & 49.2 & 53.6 & 58.6 & - & 61.2 \\ 
			Pretrained  & 14.9 & 34.5 & 53.1 & 54.4 & 60.0 & 61.0 & 61.1 \\ 
			MoCo-v3-512 (ours) & 	\textbf{50.4} & \textbf{53.1} & \textbf{55.2} & 54.6 & \textbf{61.4} & 61.8 & 62.4
			\\ 
			MoCo-v3-1024 (ours) & 49.5 & 52.2 & 52.9 & \textbf{55.3} & 61.3 & 61.9 & 62.4 \\
			MoCo-v3-4096 (ours) & 45.9 & 49.0 & 53.1 & 55.2 & 60.9 & \textbf{62.2} & \textbf{62.6} \\ 
			\textcolor{black}{SimCLR-5TX} & \textcolor{black}{39.1} & \textcolor{black}{42.4} & \textcolor{black}{52.3} & \textcolor{black}{53.3} & \textcolor{black}{57.9} & \textcolor{black}{60.3} & \textcolor{black}{60.9} \\ 
			\textcolor{black}{SimCLR-AWGN \cite{SSL_Yun}} & \textcolor{black}{26.9} & \textcolor{black}{34.8} & \textcolor{black}{50.2} & \textcolor{black}{53.1} & \textcolor{black}{57.7} & \textcolor{black}{60.5} & \textcolor{black}{60.4}
			\\
			MCLDNN \cite{MCLDNN} & - & - & - & - & - & 62.1 & - \\
			PET-CGDNN \cite{amr_benchmark} & - & - & - & - & - & 60.4 & - \\ \hline
			\# Training samples &  880  & 1760 & 8.8K & 17.6K & 88K & 132K &  158.4K \\
			\hline
	\end{tabular}}
\end{table}

\textcolor{black}{Table~\ref{tab:amr_comprasion} studies the effect of different initializations}. We observe that the Xavier initialized model performs the worst across all training and validation splits. The proposed MoCo-v3 pretrained model achieves the best performance in all splits. When only 0.5\% of training data is used over all SNR values, the Xavier initialized model and the pretrained model achieve only 9.1\% and 14.9\% accuracy (note that since there are 11 labels, purely random classification would achieve $\frac{100}{11}$ $\approx$ 9.1\% accuracy), whereas the proposed model with projection size of 512 (MoCo-v3-512) achieves 50.4\% accuracy demonstrating a 35-41\% improvement. The pretrained model achieves 53.1\% only when 5\% of the data used. This indicates a 10$\times$ improvement in sample efficiency which means it takes 10 times more labeled samples for the pretrained model to achieve the same performance of MoCo-v3 pretrained model in the low-labeled data regime. 
As the number of labeled samples increases, the accuracy gap between the pretrained and self-supervised models decreases. 
The accuracy difference between the pretrained model and MoCo-v3-512 starts at ~35\% at 0.5\% data and drops to 1\% improvement when 90\% of the data is used, since the importance of pretraining diminishes as the model is trained with more labels. \rev{When we compare the performance of different SSL methods, MoCo-v3 significantly outperforms the SimCLR variants. For example, SimCLR-AWGN and SimCLR-5TX achieve 26.9\% and 39.1\% accuracy at 0.5\% data, respectively, while MoCo-v3-512 achieves 50.4\% indicating a significant difference between the SimCLR and proposed algorithms. This result also indicates the importance of transformations such that the SimCLR-5TX model consistently performs better than the SimCLR-AWGN model.}

Table~\ref{tab:ssl_projection_head} shows the effects of projection head size for different training and validation splits. Projection head uses the features extracted by the backbone CNN and improves its representation quality \cite{SimCLRv2}. The projection head size determines the width of this MLP.  The models are pretrained using different projection head sizes ranging from 128 to 4096. After pretraining, the models are end-to-end trained (with labels) using different train and validation splits. \textcolor{black}{We observe that smaller projection head sizes generally perform better in the low-data regime, while larger projection heads have better performance for high-data regime. This is probably due to the fact that a large projection head has more trainable parameters, which requires more samples to be trained on and in the low-data regime, there are not simply enough samples}.

\begin{figure}[tbh!]
	\centering
	\includegraphics[width=0.9225\columnwidth]{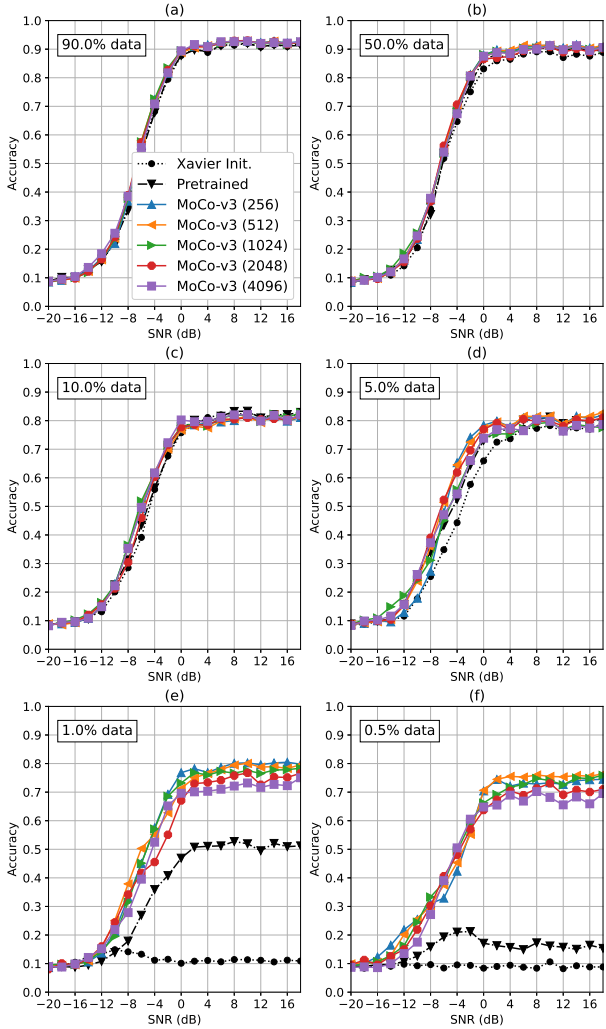}
	\caption{Accuracy vs. SNR for different training and validation splits.}
	\label{fig:performance_plots}
\end{figure}

Figs.~\ref{fig:performance_plots}(a)-(f) show the AMR accuracy (averaged over all labels) versus SNR for different splits of training and validation data. For all splits, we observe that as the SNR decreases below -4~dB, all models suffer in making accurate classifications. This is an expected result since the noise power overwhelms the signal strength and the accuracy decreases.  When 0.5\% data is used, the self-supervised classifier achieves more than 60\% accuracy for SNR$>$$-2$~dB whereas the pretrained model achieves at most 20\%. Similarly, when 1\% of the labeled data is used, the Xavier initialized and pretrained models achieve at least 9\% and 50\% accuracy at SNR$>$$2$~dB, respectively, while all MoCo-v3 pretrained models achieve more than 70\% accuracy. As the number of labeled data increases, all models perform very similar to each other. However, self-supervised representation learning still provides advantages. As an example, when 90\% of the data is used,  the pretrained model achieves 91.3\% accuracy at 18~dB SNR, whereas the MoCo-v3-512 model achieves 92.4\%. 

\begin{table}[tb!]
	\small
	\centering
	\caption{\rev{Effect of projection head size in MoCo-v3 on the AMR accuracy.}}
	\label{tab:ssl_projection_head}
	\resizebox{\columnwidth}{!}{
		\begin{tabular}{c|ccccccccc}
			\hline	
			Projection Size & 0.5\% & 1\% & 5\%    & 10\%   & 50\% &  75\% & 90\% \\
			\hline
			256 & 49.5 & \textbf{53.5} & 54.2 & 54.6 & 61.2 & 61.3 & 62.3 \\ 
			512 & \textbf{50.4} & 53.1 & \textbf{55.2} & 54.6 & \textbf{61.4} & 61.8 & 62.4 \\ 
			1024 & 49.5 & 52.2 & 52.9 & \textbf{55.3} & 61.3 & 61.9 & 62.4 \\ 
			2048 & 47.6 & 49.8 & 54.7 & 54.2 & 60.7 & 61.7 & 62.4\\
			4096 & 45.9 & 49.0 & 53.1 & 55.2 & 60.9 & \textbf{62.2} & \textbf{62.6} \\ 
			\hline
	\end{tabular}}
\end{table}

\textcolor{black}{As a final note, we compare the size and computational complexity of the models. The ResNet-50 model has 23.52M parameters, whereas MCLDNN and PETCGDNN have 406.2K and 71.9K, respectively. The computational complexity of the ResNet-50 model is 293.4M floating-point operations (flops), which is substantially higher than the MCLDNN with 0.7M flops and PETCGDNN with 1.3M flops. Despite the additional complexity of ResNet-50, we note that our main contribution is to use the SSL method to improve the AMR performance rather than developing a new architecture.
}

\section{Conclusion} \label{sec:Conclusion}
In this letter, we studied the AMR problem using the DL models. The AMR models using SSL can learn signal representations without any annotation. This property can be leveraged to increase the model performance and sample efficiency. For this goal, we proposed a contrastive learning algorithm using the MoCo-v3 approach. In addition, we proposed five data augmentation methods to learn the signal representations. The learned representations increase the sample efficiency, which means that the model can be fine-tuned with less labeled data to achieve high performance. On the other hand, when there \rev{is} a lot of labeled data, the learned representations are still able to improve the classification accuracy. \rev{In future work, we plan to evaluate the performance of other SSL algorithms and additional signal transformations. In addition, we plan to extend the self-supervised signal representation learning framework to other RF tasks such as emitter identification and radar waveform recognition.} 
\bibliographystyle{IEEEtran}
\bibliography{refs2}

\end{document}